\documentclass{article}

\usepackage{microtype}
\usepackage{graphicx}
\usepackage{subfigure}
\usepackage{booktabs} 
\usepackage{multirow}

\usepackage{hyperref}

\usepackage[accepted]{icml2024}

\usepackage{amsmath}
\usepackage{amssymb}
\usepackage{mathtools}
\usepackage{amsthm}

\usepackage[capitalize,noabbrev]{cleveref}

\theoremstyle{plain}

\theoremstyle{definition}

\theoremstyle{remark}

\usepackage[textsize=tiny]{todonotes}

\icmltitlerunning{RNAFlow: RNA Structure \& Sequence Design via Inverse Folding-Based Flow Matching}

\begin{document}

\twocolumn[
\icmltitle{RNAFlow: RNA Structure \& Sequence Design via \\ Inverse Folding-Based Flow Matching}



\icmlsetsymbol{equal}{*}

\begin{icmlauthorlist}
\icmlauthor{Divya Nori}{yyy}
\icmlauthor{Wengong Jin}{xxx}
\end{icmlauthorlist}

\icmlaffiliation{yyy}{Department of Electrical Engineering and Computer Science, Massachusetts Institute of Technology, Cambridge, MA, USA}
\icmlaffiliation{xxx}{Broad Institute of MIT and Harvard, Cambridge, MA, USA}

\icmlcorrespondingauthor{Divya Nori}{divnor80@mit.edu}

\icmlkeywords{Machine Learning, ICML}

\vskip 0.3in
]



\printAffiliationsAndNotice{} 

\begin{abstract}
The growing significance of RNA engineering in diverse biological applications has spurred interest in developing AI methods for structure-based RNA design. While diffusion models have excelled in protein design, adapting them for RNA presents new challenges due to RNA's conformational flexibility and the computational cost of fine-tuning large structure prediction models. To this end, we propose RNAFlow, a flow matching model for protein-conditioned RNA sequence-structure design. Its denoising network integrates an RNA inverse folding model and a pre-trained RosettaFold2NA network for generation of RNA sequences and structures. The integration of inverse folding in the structure denoising process allows us to simplify training by fixing the structure prediction network. We further enhance the inverse folding model by conditioning it on inferred conformational ensembles to model dynamic RNA conformations. Evaluation on protein-conditioned RNA structure and sequence generation tasks demonstrates RNAFlow's advantage over existing RNA design methods.
\end{abstract}

\section{Introduction}

\begin{figure*}[ht]
    \centering
    \includegraphics[width=2\columnwidth]{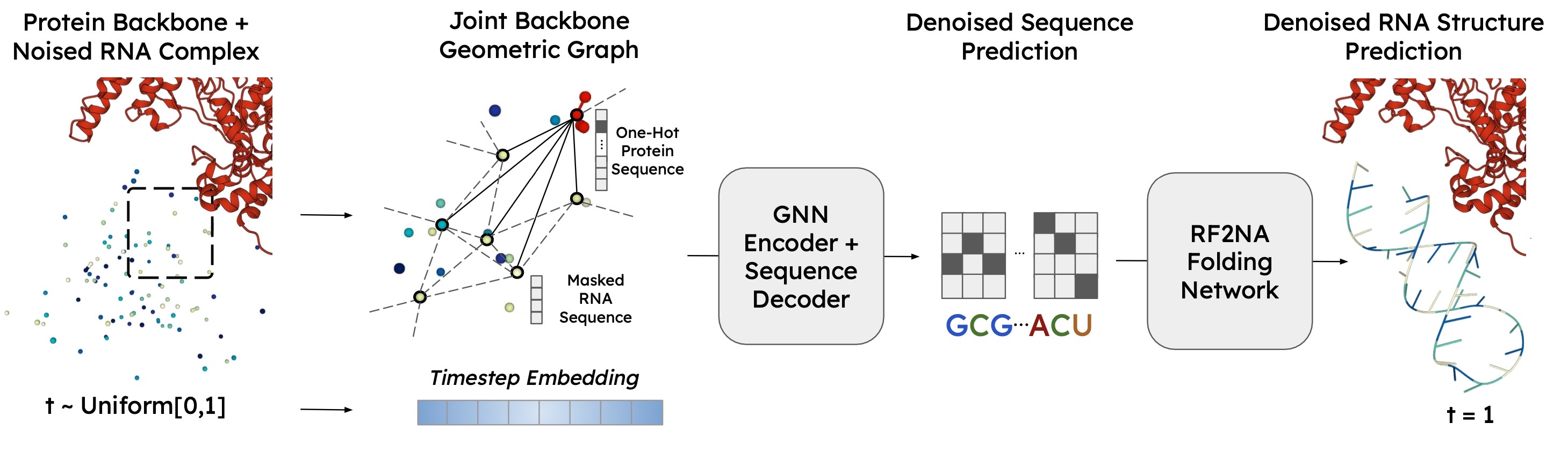}
    \caption{One forward pass in RNAFlow training, during which an inverse folding model is finetuned to be the flow matching score prediction network. The inverse folding model predicts a denoised RNA sequence from a noised complex backbone graph. The predicted sequence is folded by RF2NA for sequence and structure supervision.}
    \label{fig:1}
\end{figure*}

In recent years, RNA molecules have been engineered for versatile and controllable functions in biological systems \cite{dykstra2022engineering}, ranging from synthetic riboswitch sensors that modulate gene expression to aptamers that bind with protein targets ~\cite{thavarajah2021rna, vezeau2023automated, keefe2010aptamers}. However, experimental methods for high-throughput RNA selection like SELEX~\cite{gold2015selex} remain time-consuming and labor-intensive. To unlock the full potential of RNA therapeutics, there is a pressing need for developing deep learning models to automatically design RNAs that bind with a protein of interest \cite{sanchez2019rna}.

Current state-of-the-art biomolecular structure design methods are based on diffusion models~\cite{ho2020denoising,yim2023se} or flow matching~\cite{lipman2022flow,bose2023se}. For example, RFDiffusion~\cite{watson2023novo} fine-tuned RoseTTAFold~\cite{baek2021accurate} to generate novel protein structures that meet complex constraints, including conditional generation of protein binders. While we can adopt a similar approach for conditional RNA structure generation by fine-tuning a protein-nucleic acid structure prediction network like RoseTTAFold2NA (RF2NA)~\cite{baek2023accurate}, there are two major caveats. First, RNAs exhibit a high degree of conformational flexibility, which is often the key to their biological function \cite{ganser2019roles}. A generative model that outputs a single static structure may bottleneck the downstream sequence generation process, since an ideal sequence should account for desired RNA dynamics. Second, fine-tuning a large structure prediction network is computationally expensive, motivating a method that applies a pre-trained model like RF2NA directly. 

In this paper, we propose RNAFlow, a flow matching model for RNA sequence-structure design. The denoising network in RNAFlow is composed of an RNA inverse folding model and a pre-trained RF2NA network. In each iteration, RNAFlow first generates a RNA sequence given a noisy protein-RNA complex and then uses RF2NA to fold into a denoised RNA structure. For computational efficiency, we train the inverse folding model to minimize the flow matching objective while keeping RF2NA fixed. RNAFlow offers three advantages over previous methods like RFDiffusion. First, RNAFlow generates an RNA sequence and its structure simultaneously. Second, it is much easier to train because we do not fine-tune a large structure prediction network. Third, our framework enables us to model the dynamic nature of RNA structures for inverse folding. Specifically, we utilize the final few structure predictions from inference trajectories as an effective approximation of an RNA conformational ensemble and enhance our inverse folding model to condition on dynamic RNA conformations.

We evaluate our method on the task of protein-conditioned RNA structure and sequence generation. RNAFlow outperforms a standard sequence-only approach and a recent diffusion model \cite{morehead2023towards} for nucleic acid sequence-structure generation in terms of native sequence recovery, RMSD, and lDDT. Additionally, we show that RNAFlow can be used in the motif-scaffolding setting to generate plausible RNA aptamers for G-protein-coupled receptor kinase 2 (GRK2), a target with known sequence motif for GRK2 binding.
\section{Related Work}

\textbf{Protein-Conditioned RNA Design.} Early methods for computational design of protein-binding RNAs involved generating a large number of RNA sequences and selecting by desired secondary structure motifs \cite{kim2010computational} or molecular dynamics \cite{zhou2015searching, buglak2020methods}. These approaches are computationally expensive and require specifying design constraints a priori, which are often unknown for new protein targets. More recently, generative sequence-based approaches based on LSTMs and VAEs have been trained on SELEX data \cite{im2019generative, iwano2022generative}. However, these methods can only be trained for one protein at a time and cannot be applied to proteins where SELEX data does not exist.

\textbf{Unconditional RNA Design.} Computational methods have been developed for RNA structure design, including classical algorithmic approaches \cite{yesselman2019computational} and generative modeling methods. In particular, \citet{morehead2023towards} recently proposed an $SE(3)$-discrete diffusion model (MMDiff) for joint generation of nucleic acid sequences and structures. While MMDiff can generate short micro-RNA molecules, it has trouble conditionally generating protein-binding RNAs and samples of longer sequence length.

Deep learning methods for inverse folding generally apply graph-based encoders for a single RNA structure \cite{tan2023hierarchical} or several conformers \cite{joshi2023multi}. We adapt an inverse folding model to accept protein information as a condition and train on a denoising objective.

\textbf{Protein Structure Design.} Deep generative methods for protein design have demonstrated that realistic protein backbones can be generated efficiently via flow matching \cite{yim2023fast, bose2023se}. Prior work has also shown that diffusion models can be conditioned on a target protein to generate functional protein binders \cite{ingraham2023illuminating, watson2023novo}. We apply a similar approach within the flow matching framework. Our work is also related to sequence-structure co-design methods where the outputs are iteratively refined \cite{jin2021iterative, stark2023harmonic}.
\section{Methods}

\textbf{Problem Formulation.} As input, RNAFlow receives the protein backbone atom structure $\vec P \in \mathbb{R}^{L_p\times3\times3}$, where $L_p$ is the number of residues, and each residue contains backbone atoms $N$, $C_\alpha$, and $C$. The protein sequence is also given as input where a single token is $p_i$, $i \in \{0, 1, 2, \ldots, L_p - 1\}$. The model is trained to predict RNA sequence with tokens $r_i$, $i \in \{0, 1, 2, \ldots, L_r - 1\}$. The model also predicts RNA backbone structure $R \in \mathbb{R}^{L_r\times3\times3}$, where $L_r$ is the number of nucleotides, and each nucleotide contains backbone atoms $P$, $C_{4}'$, and $N_{1}/N_{9}$ (pyrimidine/purine)~\citep{wadley2007evaluating}.

\textbf{Background: Conditional Flow Matching.} By the flow matching framework~\cite{lipman2022flow}, consider data distribution $p_1$ of RNA backbone structures and a prior distribution $p_0(\vec R | \vec R_1)$, where $\vec R_1$ is a sample from $p_1$. A \textit{flow} transforms $p_0$ to $p_1$, and this flow can be parameterized by a sequence of time-dependent conditional probability paths $p_t(\vec R_t | \vec R_1)$, $t \in [0,1]$. A sample from $p_t$ can be computed by linear interpolation, as given by Equation \hyperref[interpolation]{1}, where $\vec R_0$ is a sample from the prior distribution.

\begin{equation}\label{interpolation}
\vec R_t | \vec R_1 = (1 - t) * \vec R_0 + t * \vec R_1
\end{equation}

This probability path is generated by a marginal vector field that defines how individual samples are transported. We can define this vector field as given in Equation \hyperref[vecfield]{2}, and we can integrate over the field from time $0$ to time $1$ to generate samples from $p_1$ given a noise sample from $p_0$.

\begin{equation}\label{vecfield}
\hat v (\vec R_t; \theta) = \frac{\hat R_1(\vec R_t; \theta) - \vec R_t}{1 - t}
\end{equation}

We aim to learn this vector field by approximating $\hat R_1$ with a neural network trained with the reparameterized conditional flow matching objective. Particularly, as with diffusion, the neural network can conveniently be trained to predict samples from the data distribution given noised sample $\vec R_t$.

We choose to employ flow matching rather than the related diffusion framework \cite{ho2020denoising}, primarily due to computational efficiency. Diffusion inference often requires thousands of forward passes to generate quality samples. Since our score model involves a large structure prediction network, RNAFlow inference time would be much slower in a diffusion setting \cite{yim2023fast}. In contrast, quality samples can be generated by flow matching with a much fewer number of passes. 

\subsection{RNAFlow Algorithm: Overview}

We generate RNA sequences and structures using a conditional flow matching model where the score predictor is an inverse folding denoiser and pre-trained folding network. The inverse folding denoiser, which we refer to as \textbf{Noise-to-Seq}, is a geometric graph neural network conditioned on protein structure and sequence. Noise-to-Seq is pre-trained on the RNA inverse folding task and subsequently fine-tuned on the flow matching objective. It follows the RNA inverse folding architecture presented by \citet{joshi2023multi}, which is detailed in section \hyperref[sec:3.2]{3.2}.

\textbf{Initialization}. As our prior distribution, we choose a unit Gaussian on $\mathbb{R}^3$ centered at zero, from which we sample $\vec R_0$. We also translate the true protein-RNA complex [$\vec P_1$, $\vec R_1$] such that the center of mass of $\vec R_1$ is zero. As shown in Algorithm \hyperref[rnaflow:train]{1}, we Kabsch align \cite{kabsch1976solution} the sampled noise $\vec R_0$ with the true RNA $\vec R_1$ to eliminate global rotations, leading to faster training convergence \cite{klein2023equivariant}. By this approach, we train with the ground-truth pose of the protein relative to the RNA centroid, though this is not leaked at inference time.

\begin{algorithm}[t]
\caption{RNAFlow: Train}\label{rnaflow:train}

\algorithmicrequire $\{p_i\}_{\forall i}$, $\{r_i\}_{\forall i}$, $[\vec P_1, \vec R_1]$
\vspace{0.25em}

Sample prior $\vec R_0 \sim \mathcal{N}(0, I_3)^{L_r}$
\vspace{0.25em}

Kabsch align noise $\vec R_0$ with true backbone $\vec R_1$
\vspace{0.25em}

Sample timestep $t \sim \text{Uniform}[0, 1]$
\vspace{0.25em}

Interpolate $\vec R_t \gets t * \vec R_1 + (1 - t) * \vec R_0$
\vspace{0.25em}

Predict $\{\hat r_i\}_{\forall i} \gets \text{\textbf{Noise-to-Seq}} \{ [\vec P_1, \vec R_t], \{p_i\}_{\forall i}, t \}$ 
\vspace{0.25em}

Predict $\hat R_1 \gets \text{RF2NA} \{ \{\hat r_i\}_{\forall i} \} $

\end{algorithm}

\textbf{Training}. During training, we sample a timestep $t$ and interpolate the true RNA backbone with the sampled prior to arrive at noised backbone $\vec R_t$. The true protein backbone, noised RNA backbone, and protein sequence are given as input to Noise-to-Seq which predicts a denoised RNA sequence. RF2NA folds the predicted RNA sequence into predicted structure $\hat R_1$. This process is shown in Figure \ref{fig:1}.

The outputs of an RNAFlow training pass are an RNA sequence and its structure. The reparameterized flow matching objective gives that the score model should be trained with a standard $L_2$ loss. Therefore, as given in Equation \hyperref[loss]{3}, we compute the MSE between all predicted and true RNA backbone coordinates. Before MSE computation, we Kabsch align the two RNA backbones, since we are optimizing for RNA structure design rather than docking accuracy. Additionally, we supervise the sequence with a cross entropy loss between predicted and true nucleotide types.

\begin{equation}\label{loss}
\mathcal{L} = \text{MSE}(\hat R_1, \vec R_1) + \sum_i \text{CE}(\hat r_i, r_i)
\end{equation}

The sequence output from Noise-to-Seq is used as input to RF2NA, and the resulting structure is supervised to update the weights of the model. Since gradients must propagate between Noise-to-Seq and RF2NA, we apply the Gumbel-Softmax estimator to differentiably sample from the predicted logits \cite{jang2016categorical}. Further justification of this objective is given in the Appendix. 

\textbf{Inference: RNAFlow-Base}. As shown in Algorithm \hyperref[rnaflow:inference]{2}, we begin by generating a ``pose guess'' using RF2NA. Specifically, we fold the protein MSA with a mock RNA sequence consisting of all adenines, of the same length as the true RNA sequence. The predicted complex is used as an initial guess of protein pose with respect to the RNA centroid. $\vec P_0$ is the true protein backbone Kabsch-aligned onto the predicted protein pose, and $\vec R_0$ is drawn from the prior distribution. RNAFlow refines its predictions over multiple steps, predicting the true complex [$\hat P_1$, $\hat R_1$] on each iteration. The final outputs of RNAFlow inference are a predicted RNA sequence and structure.

\textbf{Inference: RNAFlow-Traj}. In standard RNAFlow inference, Noise-to-Seq predicts a sequence based on a single RNA structure. However, an RNA backbone structure can adopt multiple conformations, and it is important that the final RNA sequence reflects this conformational diversity \cite{ganser2019roles}. Indeed, \citet{joshi2023multi} demonstrates that inverse folding on a set of RNA conformations improves the native sequence recovery rate compared to single-structure model.
To this end, we propose to generate a final RNA sequence by conditioning on multiple RNA structures generated over the course of flow matching inference. This inverse folding model, referred to as \textbf{Trajectory-to-Seq (Traj-to-Seq)}, is a multi-graph neural network which can handle multiple RNA conformation inputs. Its architecture is detailed in section \hyperref[sec:3.3]{3.3}.

\begin{algorithm}[h]
\caption{RNAFlow-Traj: Inference}\label{rnaflow:inference}

\algorithmicrequire $\{p_i\}_{\forall i}$, $\vec P_1$
\vspace{0.25em}

Position $\vec P_0$ by aligning $\vec P_1$ with RF2NA pose guess
\vspace{0.25em}

Sample prior  $\vec R_0 \sim \mathcal{N}(0, I_3)^{L_r}$
\vspace{0.25em}

Initialize $traj \gets []$
\vspace{0.25em}

\algorithmicfor {$n \gets 1$ \textbf{to} $N$}
\vspace{0.25em}

    \hspace{0.2in} Let $t_2 \gets n/N$ and $t_1 \gets (n-1)/N$
    \vspace{0.25em}
    
    \hspace{0.2in} $\{\hat r_i\}_{\forall i} \gets \text{\textbf{Noise-to-Seq}} \{ [\vec P_{t_1}, \vec R_{t_1}], \{p_i\}_{\forall i}, t_1 \}$
    \vspace{0.25em}
    
    \hspace{0.2in} Predict [$\hat P_1$, $\hat R_1] \gets \text{RF2NA} \{ \{\hat r_i\}_{\forall i}, \text{protein MSA} \}$
    \vspace{0.25em}
    
    \hspace{0.2in} Append $\hat R_1$ to $traj$
    \vspace{0.25em}
    
    \hspace{0.2in} Interpolate $\vec R_{t_2} \gets \vec R_{t_1} +  \frac{(\hat R_{1} - \vec R_{t_1})}{(1 - t_1)} * (t_2 - t_1)$
    
    \hspace{0.2in} Align $\vec P_{t_2} \gets \text{Kabsch}(\vec P_1, \hat P_1)$
    
\algorithmicendfor

Predict $\{\hat r_i\}_{\forall i} \gets \text{\textbf{Traj-to-Seq}} \{ traj \}$
\vspace{0.25em}

\textbf{Outputs:} $\{\hat r_i\}_{\forall i}$, $traj[-1]$

\end{algorithm}

\subsection{Noise-to-Seq Module}
\label{sec:3.2}

Noise-to-Seq is a graph-based RNA inverse folding model, fine-tuned to predict RNA sequences autoregressively from noised structures. As in \citet{ingraham2019generative} and \citet{jing2021equivariant}, the model applies an encoder-decoder architecture where the encoder learns a representation of the protein-RNA complex structure and the decoder predicts a distribution over nucleotides at each position $i$, given sequence information of all positions before $i$.

\textbf{Graph Representation.} We represent each backbone 3D point cloud [$\vec P$, $\vec R$] as a graph $\mathcal{G} = (\mathcal{V}, \mathcal{E})$. Each amino acid $i$ is assigned a node at $C_\alpha$ coordinate $\vec{x_i} \in \vec P$, and each nucleotide $j$ is assigned a node at $C_{4}'$ coordinate $\vec{x_j} \in \vec R$. Node features are denoted as $\mathcal{V} = \{ v_1, \ldots, v_{L_p + L_r} \}$. As shown in Figure \hyperref[fig:1]{1}, for all nodes $\vec{x}_i \in \vec P$, we draw edges to $10$ nearest neighbor nodes whose 3D coordinates are also in $\vec P$. Likewise for all nodes $\vec{x}_i \in \vec R$, we draw edges to $10$ nearest neighbor nodes whose 3D coordinates are also in $\vec R$. For all nodes $\vec{x}_i \in \vec P$, we additionally draw edges to $5$ nearest neighbor nodes $\vec{x}_j \in \vec R$. Nearest neighbors are computed by $\Vert \vec{x}_i - \vec{x}_j \Vert_2$. Edge features are denoted as $\mathcal{E} = \{ e_{i,j} \}_{i \neq j}$.

\textbf{Node \& Edge Features}. We compute node features and edge features as follows. Node vector features include unit vectors to neighboring nodes on the backbone chain and unit vectors to the remaining two atoms of the nucleotide ($P$ and $N_{1}/N_{9}$) or amino acid ($N$ and $C$). Node scalar features include magnitudes of all vector features, encoded by radial basis functions, and a one hot encoding of residue identity for the protein chain. Edge vector features include a unit vector from the source node to the destination node, and its magnitude encoded by radial basis functions is a scalar feature. Edge scalar features also include distance along the backbone between the source and destination, encoded with a sinusoidal embedding. If the edge connects nodes in two different chains, the encoded number is equal to $L_p+L_r$.

\begin{figure*}[ht]
    \centering
    \includegraphics[width=2\columnwidth]{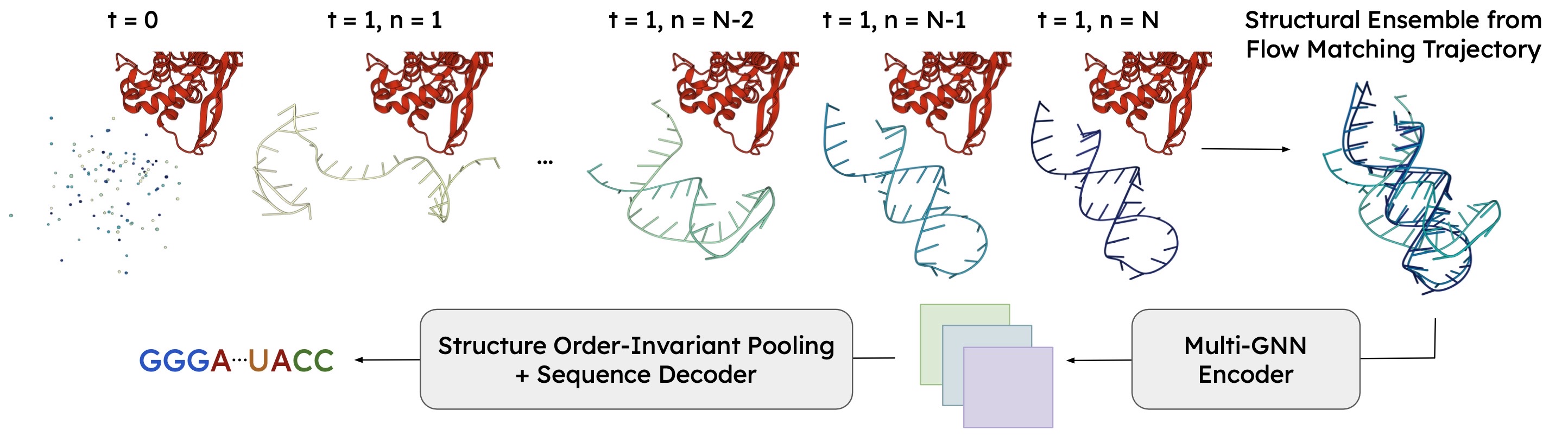}
    \caption{One forward pass during Traj-to-Seq inference. A subset of structures from a flow matching trajectory are encoded by a multi-GNN and pooled in an order-invariant manner to predict an RNA sequence.}
    \label{fig:2}
\end{figure*}

\textbf{Model Architecture.} Noise-to-Seq contains a message-passing encoder consisting of GVP-GNN layers \cite{jing2020learning}. First, input node features $v_i$ and edge features $\{ e_{i,j} \}_{i \neq j}$ are encoded by GVPs.
\begin{align*}
h_{v_i} &= g_v ( \text{LayerNorm} (v_i) )  \\
h_{e_{i,j}} &= g_e ( \text{LayerNorm} (e_{i,j}) ) 
\end{align*}
Node embedding GVP $g_v$ and edge embedding GVP $g_e$ apply the vector gating strategy proposed by \citet{jing2021equivariant}. If the input structure is noised, we add a timestep embedding to the node features, encoded by random Fourier features of size $256$ \cite{tancik2020fourier} and concatenated onto the scalar component of $h_{v_i}$. For every position $i$, $h_{v_i}$ is residually updated by a sequence of three message-passing layers where each layer has the following architecture:
\begin{align*}
m_{v_i} &= \frac{1}{|\mathcal{N}|} \sum_{j \in \mathcal{N}} g_{MSG} (h_{v_i}, h_{v_j}, h_{e_{i,j}})  \\
h_{v_i}^{'} &= \text{LayerNorm} (h_{v_i} + \text{Dropout}(m_{v_i})) \label{update}
\end{align*}
$g_{MSG}$ is a sequence of three GVPs with vector gating and ReLU activation on the scalar features. After each message-passing layer, we also apply pointwise feedforward updates to the node embedding.
\begin{align*}
h_{v_i}^{'} = \text{LayerNorm} (h_{v_i}^{'} + \text{Dropout}(g_{FF}(h_{v_i}^{'})))
\end{align*}
$g_{FF}$ is a sequence of two GVPs with vector gating and the ReLU activation on the scalar features. We then encode protein and RNA sequence information with an embedding layer and concatenate these features onto the edge scalar features. By the autoregressive scheme, we mask out sequence features for edges where the source node's position is greater than the destination node's position. The protein nodes are always positioned before RNA nodes, so protein sequence context is always present during decoding. Our decoder's architecture is identical to the encoder, except that messages are aggregated by a sum instead of a mean. Finally, we apply a GVP layer and softmax to predict probabilities of the $4$ nucleotide classes at each position. The model is supervised by a cross entropy loss between the true RNA sequence and predicted nucleotide probabilities.

\subsection{Traj-to-Seq Module}
\label{sec:3.3}

As shown in Figure \hyperref[fig:2]{2}, the flow matching inference loop generates a trajectory of RNA structures over iterative refinement steps. Traj-to-Seq is a graph-based inverse folding model that predicts RNA sequences based on the output trajectory of a single RNAFlow inference pass. At the final stages of the trajectory, the structures have a high degree of secondary structure similarity while displaying variation in tertiary structure, qualitatively resembling a conformational ensemble \cite{fornili2013specialized}. Therefore, we can leverage the outputs of a single flow matching inference pass as a conformational ensemble approximation.

\textbf{Model Inputs.} We represent a trajectory of RNA backbones as a set of independent graphs  \{$\mathcal{G}^{(1)}, ..., \mathcal{G}^{(k)}$\} where $k$ is the number of conformers. Traj-to-Seq does not accept protein information as input, so each node in $\mathcal{G}^{(n)}$ corresponds to a nucleotide $j$ at $C_{4}'$ coordinate $\vec{x_j} \in \vec R^{(n)}$. For all nodes $\vec{x}_j \in \vec R^{(n)}$, we draw edges to $10$ nearest neighbor nodes whose 3D coordinates are also in $\vec R^{(n)}$, so edges are not drawn between conformers. Instead, Traj-to-Seq is designed to operate on ``multi-graphs.'' As introduced by \citet{joshi2023multi}, a multi-graph is constructed by stacking each RNA graph's scalar and vector features and building a joint adjacency matrix, computed as the union of each individual graph's adjacency matrix.

\textbf{Architecture.} Traj-to-Seq employs the same encoder-decoder architecture as Noise-to-Seq. The GVP encoder processes each structure independently, such that $h_{v_i} \in \mathbb{R}^{k \times f}$ where $k$ is the number of conformers and $f$ is the feature dimension. A merged representation is computed by a structure order-invariant mean. Finally, a decoder identical to Noise-to-Seq is employed to predict RNA sequence. 

\subsection{Output Rescoring Model}

Since RNAFlow can be sampled many times to obtain several RNA designs, we train an \textit{output rescoring model} to select from amongst the samples based on predicted recovery rate. For training, we generate 6 mutated RNA sequences for each ground-truth protein-RNA complex. Binary labels are generated based on whether or not a generated RNA sequence has a recovery rate $\geq 30\%$. At test time, we evaluate many RNAFlow output samples for a single protein and make selections based on the highest predicted probability of a positive outcome, prioritizing designs with the greatest likelihood of achieving a recovery rate of at least $30\%$.

The inputs to the output scoring model are protein sequence, protein structure, RNA sequence, and RF2NA-folded RNA structure. The protein and RNA are each encoded by a GVP model with the same architecture as the Noise-to-Seq encoder, and the resulting node-level representations are averaged and processed by a feed-forward network. The output is supervised by a binary cross entropy loss. Hyperparameter and training details are included in the Appendix.

\section{Experiments}

\begin{table*}[t!]
    \centering
    \small
    \caption{Structure generation results. We report Mean $\pm$ SEM for RMSD and lDDT. RMSD is computed after Kabsch alignment with the true RNA. lDDT is computed on $C_{\alpha}$ atoms.}
    \vspace{0.05in}
    \begin{tabular}{lccc@{\hskip 0.25in}ccc}
        \hline
        \multirow{2}{*}{Method} &
        \multicolumn{2}{c}{\textit{RF2NA Pre-Training Split}} &
        \multicolumn{2}{c}{\textit{Sequence Similarity Split}} \\
        & RMSD & lDDT & RMSD & lDDT \\ \hline
        Conditional MMDiff & $14.82 \pm 1.01$ & $0.34 \pm 0.02$ & $17.42 \pm 0.86$ & $0.38 \pm 0.01$ \\ \hline
        RNAFlow-Base & $12.85 \pm 0.63$ & $0.51 \pm 0.01$ & $14.77 \pm 0.34$ & \textbf{0.57} $\pm$ \textbf{0.01} \\
        RNAFlow-Traj & $13.12 \pm 0.64$ & 0.52 $\pm$ 0.01 & $15.11 \pm 0.33$ & $0.57 \pm 0.00$ \\
        RNAFlow-Base + Rescore & \textbf{10.61} $\pm$ \textbf{1.73} & \textbf{0.53} $\pm$ \textbf{0.03} & \textbf{14.60} $\pm$ \textbf{1.05} & $0.56 \pm 0.02$ \\
        RNAFlow-Traj + Rescore & $15.30 \pm 1.89$ & $0.52 \pm 0.03$ & $15.31 \pm 0.93$ & $0.56 \pm 0.02$ \\
         \hline
        RF2NA [Upper Bound] & $4.67 \pm 1.29$ & $0.76 \pm 0.04$ & $9.83 \pm 1.69$ & $0.79 \pm 0.02$ \\ \hline
    \end{tabular}
    
    \label{tab:1}
\end{table*}

\begin{table*}[h]
    \centering
    \small
    \caption{Sequence generation results. We report Mean $\pm$ SEM for native sequence recovery.}
    \vspace{0.05in}
    \begin{tabular}{lcc@{\hskip 0.25in}cc}
        \hline
        \multirow{2}{*}{Method} &
        \multicolumn{1}{c}{\textit{RF2NA Pre-Training Split}} &
        \multicolumn{1}{c}{\textit{Sequence Similarity Split}} \\
        & Recovery  & Recovery \\ \hline
        Random & $0.25 \pm 0.00$ & $0.25 \pm 0.00$ \\
        LSTM & $0.27 \pm 0.01$ & $0.24 \pm 0.01$ \\
        Conditional MMDiff & $0.24 \pm 0.02$ & $0.22 \pm 0.02$ \\ \hline
        RNAFlow-Base & $0.30 \pm 0.02$ & $0.30 \pm 0.01$ \\ 
        RNAFlow-Traj & $0.31 \pm 0.01$ & $0.28 \pm 0.01$ \\ 
        RNAFlow-Base + Rescoring & $0.33 \pm 0.02$ & \textbf{0.32} $\pm$ \textbf{0.03} \\
        RNAFlow-Traj + Rescoring & \textbf{0.37} $\pm$ \textbf{0.05} & $0.29 \pm 0.02$ \\ \hline
        Inverse Folding [Upper Bound] & $0.46 \pm 0.01$ & $0.35 \pm 0.01$ \\ \hline
    \end{tabular}
    
    \label{tab:2}
\end{table*}

We evaluate RNAFlow in two experimental settings. First, we evaluate RNA sequence and structure prediction error with respect to existing protein-RNA complexes. Dataset construction and results are described in Section \hyperref[sec:exp1]{4.1}. We additionally evaluate the novelty of RNA designs in Section \hyperref[sec:4.2]{4.2}. In Section \hyperref[sec:4.3]{4.3}, we perform an ablation study to determine how modeling choices affect structure and sequence generation.

Second, we evaluate whether RNAFlow can design RNA aptamers for the GRK2 protein, a target of interest for the treatment of chronic heart failure \cite{tesmer2012molecular}. \citet{mayer2008rna} discovered an RNA aptamer that interacts strongly with the GRK2 kinase domain, and \citet{lennarz2015rna} identified $4$ nucleotides that are critical for binding. In Section \hyperref[sec:4.4]{4.4}, we investigate whether RNAFlow can design realistic RNAs for GRK2 binding given this binding site sequence motif.

\subsection{Structure \& Sequence Accuracy}
\label{sec:exp1}

\begin{figure*}[ht]
    \centering
    \includegraphics[width=2\columnwidth]{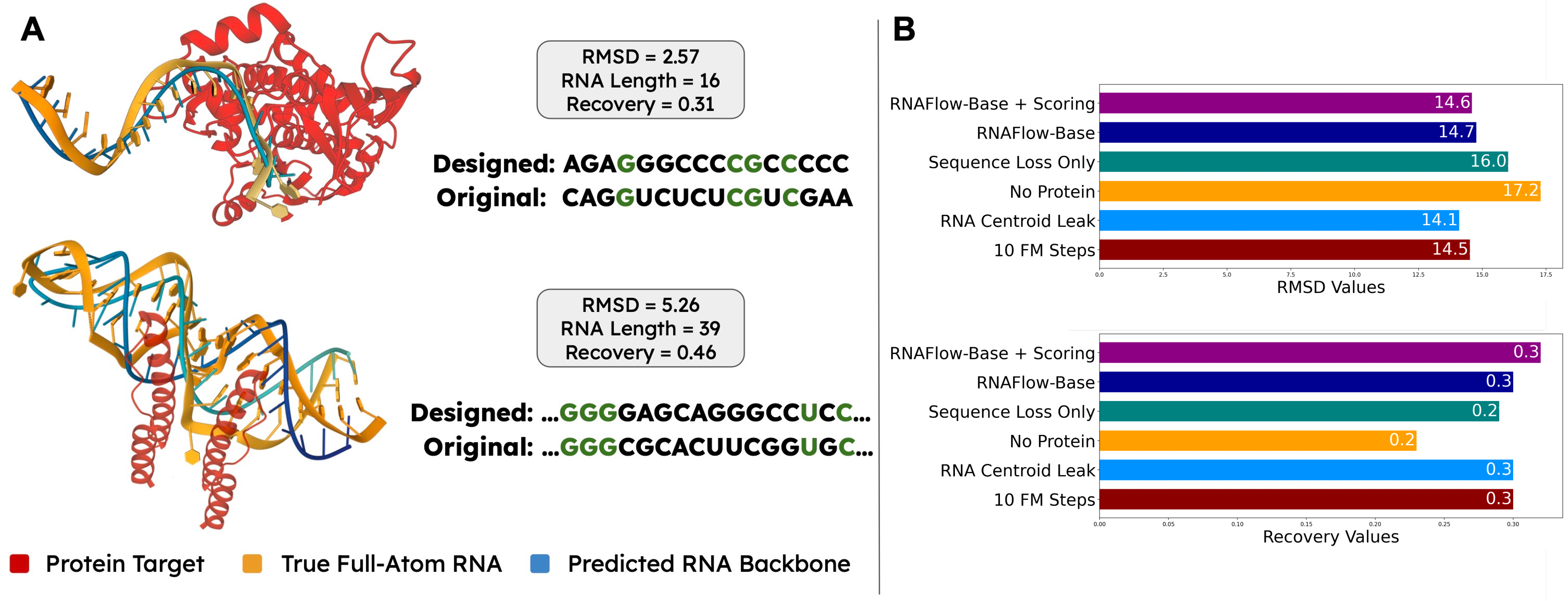}
    \caption{(A) Top: Structure and sequence design of RNA for interaction with a viral RNA-dependent RNA polymerase (PDB ID: 4K4X). Bottom: Design of RNA for interaction with HIV-1 Rev protein (PDB ID: 4PMI). (B) Ablation study of RNAFlow components. We report RMSD and sequence recovery on the sequence similarity split.}
    \label{fig:3}
\end{figure*}

As mentioned, we evaluate two variants of our method: \textit{RNAFlow-Base} which is based on Noise-to-Seq and \textit{RNAFlow-Trajectory} which additionally applies Traj-to-Seq. For both methods, we begin by generating $10$ structure-sequence pairs per protein in the test set. From amongst the $10$ designs, we additionally report the performance when $1$ sample is selected by the output rescoring model. All samples are generated with $5$ integration steps.

\textbf{Dataset.} Protein-RNA complexes from the PDBBind dataset (2020 version) were used for training and evaluation \cite{liu2017forging}. Dataset preparation details are described in the Appendix. We perform all experiments on two splits. The first accounts for RF2NA pre-training -- all examples from complexes in the RF2NA validation or test sets were assigned to the test split, and remaining examples were randomly split into training and validation in a $9$:$1$ ratio. To measure how well our model can generalize to dissimilar RNAs, we also evaluate on an RNA sequence similarity split. All RNA chains were clustered using CD-HIT \cite{fu2012cd} where two sequences are in the same cluster if they share $\geq 80\%$ sequence identity, following the clustering protocol of \citet{joshi2023multi}. The clusters were randomly split into train, validation, and test in an $8$:$1$:$1$ ratio. We pre-train Noise-to-Seq on clean protein-RNA complexes before fine-tuning on noised complexes as part of the RNAFlow training loop. Further details are included in the Appendix.

Traj-to-Seq was trained separately on RNASolo \cite{adamczyk2022rnasolo}, a dataset of RNA sequences and structures where many of the sequences are associated with multiple structure conformers. On average, the dataset contains $3$ structures per sequence. Following \citet{joshi2023multi}, we filtered to structures with resolution $\leq 3$ \AA. For the RF2NA split, we curate the Traj-to-Seq training set such that RNAs in our validation and test sets are not used for training. For the sequence similarity split, we remove RNAs that have $\geq 80\%$ similarity with data points in validation/test.  Traj-to-Seq inference was run on the final $3$ structures in the flow matching trajectory which yields optimal performance.

\textbf{Baselines.} For structure generation, we compare our method to two baselines. To our knowledge, the only existing structure generative model for nucleic acids is MMDiff \cite{morehead2023towards}, an $SE(3)$-discrete diffusion model that jointly designs nucleic acid sequences and structures. We sample $10$ RNA designs of the ground-truth RNA's length per protein-RNA complex in our test set, conditioned on the protein backbone structure and sequence. As an upper bound, we compare against RNA structures folded by RF2NA, given the ground-truth protein MSA and RNA sequence. This baseline captures the structure prediction error inherent to our score prediction model.

For sequence generation, we compare our method to four baselines. First, we compare to a random baseline where RNA sequences of specified length are generated by randomly selecting a nucleotide for each position, from a uniform distribution over \{A, C, G, U\}. Next, we compare against sequences designed by MMDiff. Our third baseline is a standard sequence-only model, consisting of an LSTM layer to encode protein sequence $\{p_i\}_{\forall i}$ and an autoregressive LSTM decoder layer to predict RNA sequence $\{\hat r_i\}_{\forall i}$, matching the architecture in \citet{im2019generative}. Finally, we establish an upper bound by comparing against results from our pre-trained inverse folding model, which takes the ground-truth protein-RNA backbone complex as input.

\textbf{Metrics.} For structure generation, we report root mean square deviation (RMSD) between predicted RNA structure and ground-truth structure, computed on all backbone atoms. Before RMSD calculation, the RNA structures are aligned by the Kabsch algorithm. For RNAFlow-Trajectory, we report the average RMSD across structures in the trajectory. We also report lDDT computed on $C_{\alpha}$ atoms. For sequence generation, we report recovery rate which is the percentage of correctly recovered nucleotides in a sampled sequence.

\textbf{Results.} As shown in Table \hyperref[tab:1]{1}, RNAFlow significantly outperforms the baseline on the structure generation task. On the RF2NA and sequence similarity splits respectively, the best RNAFlow model gives a $28\%$ and $16\%$ reduction in RMSD relative to MMDiff. We also observe a $56\%$ and $50\%$ increase in lDDT. Without rescoring, RNAFlow still gives a $13\%$ and $15\%$ reduction in RMSD. However, there is a large gap between the structural accuracy of RNAFlow and RF2NA, given the complexity of the \textit{de novo} design task in comparison to folding. We note that RNAFlow-Trajectory consistently has a higher RMSD than RNAFlow-Base, likely because trajectory structures are taken from earlier points in flow matching inference. We show two examples of generated outputs in Figure \hyperref[fig:3]{3a}, where it is evident that the designed RNA matches the ground truth structure well.

As shown in Table \hyperref[tab:2]{2}, RNAFlow outperforms all baselines on the sequence generation task. On the RF2NA split, the best RNAFlow model gives a $48\%$ improvement over random generation and a $37\%$ improvement over the LSTM baseline. RNAFlow-Trajectory outperforms RNAFlow-Base, suggesting that flow matching trajectories provide useful RNA dynamics information that is not present in a single structure. The output rescoring model improves the mean recovery, indicating that while RNAFlow samples can be variable, we learn to select quality samples. Without rescoring, we see a $15\%$ improvement in recovery relative to the best baseline.

\begin{figure*}[ht]
    \centering
    \includegraphics[width=2\columnwidth]{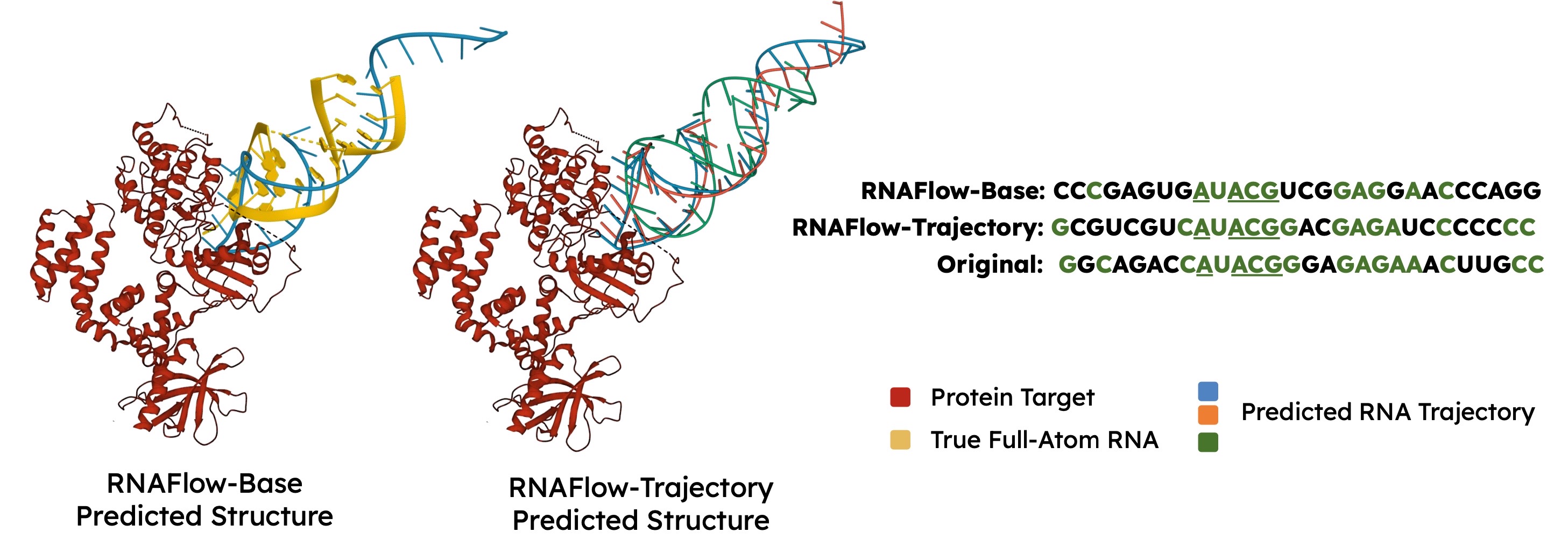}
    \caption{Structure and sequence RNAs designed by RNAFlow for GRK2 binding in motif-scaffolded setting. The predicted structure is Kabsch aligned onto the ground-truth for visualization. In the sequence designs, green-colored characters show nucleotides that are correctly recovered from the ground-truth. Underlined nucleotides are part of the given binding motif.}
    \label{fig:4}
\end{figure*}

On the sequence similarity split, the best RNAFlow model gives a $28\%$ improvement over random generation and a $33\%$ improvement over LSTM. Without rescoring, we see a $20\%$ improvement in recovery relative to the best baseline. We find that while the LSTM's performance degrades on the sequence similarity split, RNAFlow's performance is consistent. RNAFlow-Trajectory does not improve recovery, indicating that the generated trajectories may not approximate conformational ensembles as effectively when trying to generalize to biologically dissimilar RNAs. While the RNAFlow recovery rate does not reach the upper bound established by the inverse folding model, we note that for some proteins, our model designs new sequences that fold into the desired structure. For example, as shown Figure \hyperref[fig:3]{3a}, RNAFlow can design molecules that match the ground-truth structure with a relatively dissimilar sequence composition.

\subsection{Structure \& Sequence Novelty}
\label{sec:4.2}

We then evaluate the novelty of RNAs designed by RNAFlow-Base. Sequence novelty is quantified as $1 -$ sequence recovery of each generated sample to its most similar RNA sequence in the training set. Similarly, structure novelty is $1 - $ TM-score between a generated sample and its most similar RNA structure in the training set. We observe that the majority of sequence content in the generated RNAs is not observed in the training set, showing that RNAFlow can generalize and design novel RNAs. The generated structures are also reasonably novel.

\begin{table}[h]
    \centering
    \small
    \caption{Structure and sequence novelty of RNAs generated by RNAFlow-Base.}
    \vspace{0.05in}
    \renewcommand\thetable{3}
    \begin{tabular} {lcc@{\hskip 0.25in}cc}
        \hline
        & Sequence Novelty & Structure Novelty \\ \hline
        RF2NA Split & $0.55$ & $0.63$ \\
       Seq Sim Split & $0.63$ & $0.64$ \\
    \hline
    \end{tabular}
    \label{tab:3}
\end{table}

\subsection{Ablation Study}
\label{sec:4.3}

We conduct an ablation study to determine which components of our model contribute to its performance. As shown in Figure \hyperref[fig:3]{3b}, we report recovery rate and RMSD on the sequence similarity split under various conditions, since the sequence split is more general to a design setting. First, we fine-tune Noise-to-Seq on sequence cross-entropy only, removing the structure MSE loss. When this denoiser is used in the standard five-step inference loop, the RMSD increases by $1.25$ compared to RNAFlow-Base, and recovery rate decreases by $1$ percentage point. Next, we remove protein conditioning and find that RMSD increases by $2.52$, and recovery drops by $7$ percentage points, showing that providing protein information is important.

We assess the effect of our ``pose guess'' complex alignment method by performing inference with the true RNA centroid and protein orientation. While the recovery is not affected, the RMSD is $0.67$ better than our method. This shows that our initial pose prediction method does not degrade performance significantly. Finally, we show the performance when inference is performed with ten flow matching steps rather than five. While RMSD is better than our method by $0.25$, recovery is not affected. Given the insignificant impact on performance, we choose to use $5$ integration steps in favor of inference time speed-up.

\subsection{Motif-Scaffolded Design of RNAs for GRK2}
\label{sec:4.4}

\textbf{Design Procedure.} For this task, we train RNAFlow on all protein-RNA examples except the ground-truth GRK2-C28 protein-aptamer complex (PDB ID: 3UZS). Comparing against an LSTM and MMDiff, we evaluate the structure and sequence accuracy of GRK2-conditioned RNA design given a $4$-nucleotide binding site sequence motif. For all methods, we sample $10$ RNA designs and select $1$ top design by the output rescoring model. We report the recovery rate with respect to the true aptamer sequence of length $28$ and the RMSD with respect to its crystal structure. 

\textbf{Results.} As shown in Table \hyperref[tab:3]{3}, RNAFlow outperforms the baselines in terms of both RMSD and recovery. Compared to MMDiff, RNAFlow-Base gives an $8.16\%$ improvement in RMSD and $34.38\%$ improvement in recovery. As shown in Figure \hyperref[fig:4]{4}, the predicted structure resembles the ground-truth aptamer closer to the binding site, though the structure deviates when further away from the protein. 

\begin{table}[h]
    \centering
    \small
    \caption{GRK2 binder metrics.}
    \vspace{0.05in}
    \renewcommand\thetable{3}
    \begin{tabular} {lcc@{\hskip 0.25in}cc}
        \hline
        & RMSD & Recovery \\ \hline
        LSTM & - & $0.29$ \\
        MMDiff & $9.19$ & $0.32$ \\ \hline
        RNAFlow-Base & $8.44$ & $0.43$ \\
        RNAFlow-Trajectory & $\textbf{7.09}$ & $\textbf{0.54}$ \\ \hline
    \end{tabular}
    \label{tab:3}
\end{table}

Moreover, RNAFlow-Trajectory gives an $22.85\%$ improvement in RMSD and a $68.75\%$ improvement in recovery. Figure \hyperref[fig:4]{4} shows the predicted structural trajectory. One of the intermediate trajectory structures shown in green has a lower RMSD ($6.42$) than the final structure. However, inverse folding the intermediate structure alone yields a significantly lower recovery ($0.44$) than applying Traj-to-Seq ($0.54)$. Crucially, we observe that the sequence predicted by RNAFlow-Trajectory recovers nucleotides on either end of the aptamer, far away from the given binding motif. As noted by \citet{tesmer2012molecular}, these stem regions are critical for the aptamer's high affinity because they ``limit the number of possible conformations for the selected RNA.'' Thus, we argue that having conformational information is important to design these regions.

\section{Conclusion}

In this paper, we present RNAFlow, the first protein-conditioned generative model for RNA structure and sequence design. We show that an inverse folding model is an effective score prediction network within the flow matching framework, enabling the design of RNAs that outperform the baselines in terms of structure and sequence accuracy. Additionally, we show that by inverse folding over an inferred conformational ensemble, we can design plausible aptamers for GRK2 binding.

While RNAFlow shows an empirical advantage over existing methods, there is still a large gap to reach the level of accuracy achieved in \textit{de novo} protein design. We observe that currently, RNAFlow tends to achieve lower RMSD and higher recovery when the ground-truth RNA is more structurally stable (i.e. more base pairing). This suggests that RNAFlow is practically useful for aptamer design. On the other hand, when the RNA is very conformationally diverse (i.e. short single-stranded RNAs), RNAFlow often does not outperform a random sequence. In this setting, a pure language approach may be desired as structural supervision is less relevant.

To improve performance on this task, one major bottleneck is accuracy and efficiency of protein-RNA folding and docking models. This would allow us to supervise docked full-atom complexes and better model protein-RNA structural interaction, which is essential for designing RNAs that rely heavily on side chains for their function (i.e. ribozymes).

Code is available at \url{https://github.com/divnori/rnaflow}.

\section*{Impact Statement}

This paper presents work whose goal is to advance machine learning applications for structural biology and drug discovery. There are many potential societal consequences of our work, including application of our model for discovery of an RNA drug candidate. We highlight that molecules designed by RNAFlow have not been tested experimentally which is critical for any biological application. 

\bibliography{main}
\bibliographystyle{icml2024}

\newpage
\appendix
\onecolumn

\section{Flow Matching on RNA Structures}

We justify the main RNAFlow objective given in Equation \hyperref[loss]{3}, following much of the theory given by \citet{jing2023alphafold}. By the standard formulation of flow matching in Euclidean space $\mathbb{R}^{3L_r}$ ($L_r$ is RNA length), we expect to supervise with the reparameterized denoising objective (MSE).

$$ \mathcal{L} = \mathbb{E}_{R_1 \sim p_1, R_t \sim p_t} [{||\hat R_1 - \vec R_1||}^2] $$

$\hat R_1$ is predicted by neural network $\hat R_1 (\vec R_t; \theta)$ which accepts noised input $\vec R_t$ and a timestep embedding. While $\hat R_1$ is composed of an inverse folding model and pre-trained folding model, the denoised sequence can simply be thought of as an intermediate representation from which we predict the clean structure.

However, $\hat R_1 (\vec R_t; \theta)$ is $SE(3)$-invariant given that RF2NA is trained with $SE(3)$-invariant FAPE loss. We are thus flow matching over the Riemannian manifold given by quotient space $\mathbb{R}^{3L_r} / SE(3)$. To flow match in this new space, we must define a conditional probability path $p_t(\vec R_t | \vec R_1)$. We can generalize linear interpolation in the Euclidean case to interpolant $\vec R_t | \vec R_1 = \psi(R_0 | R_1)$ defined as the geodesic path from $R_0$ to $R_1$, where the geodesic path is Kabsch alignment followed by linear interpolation in $\mathbb{R}^{3L_r}$.

$$ R_0 = \text{Kabsch}(R_0, R_1) $$

$$ \psi(R_0 | R_1) = (1-t) *R_0 + t * R_1 $$

We define a distance metric $d$ on the manifold correspondingly, where $d$ is a valid metric because it is defined for all points on the manifold.

$$ d = ||\hat R_1 - \text{Kabsch}(\vec R_1, \hat R_1)|| $$

We can thus plug this expression into our original flow matching objective,

$$ \mathcal{L} = \mathbb{E}_{\vec R_1 | \vec R_t} [{||\hat R_1 - \text{Kabsch}(\vec R_1, \hat R_1)||}^2] = \text{MSE}(\hat R_1, \text{Kabsch}(\vec R_1, \hat R_1)) $$

which is equivalent to the structure loss expression in Equation \hyperref[loss]{3}. The cross entropy term simply exerts auxiliary sequence supervision.

\begin{figure*}[t!]
    \centering
    \includegraphics[width=0.6\columnwidth]{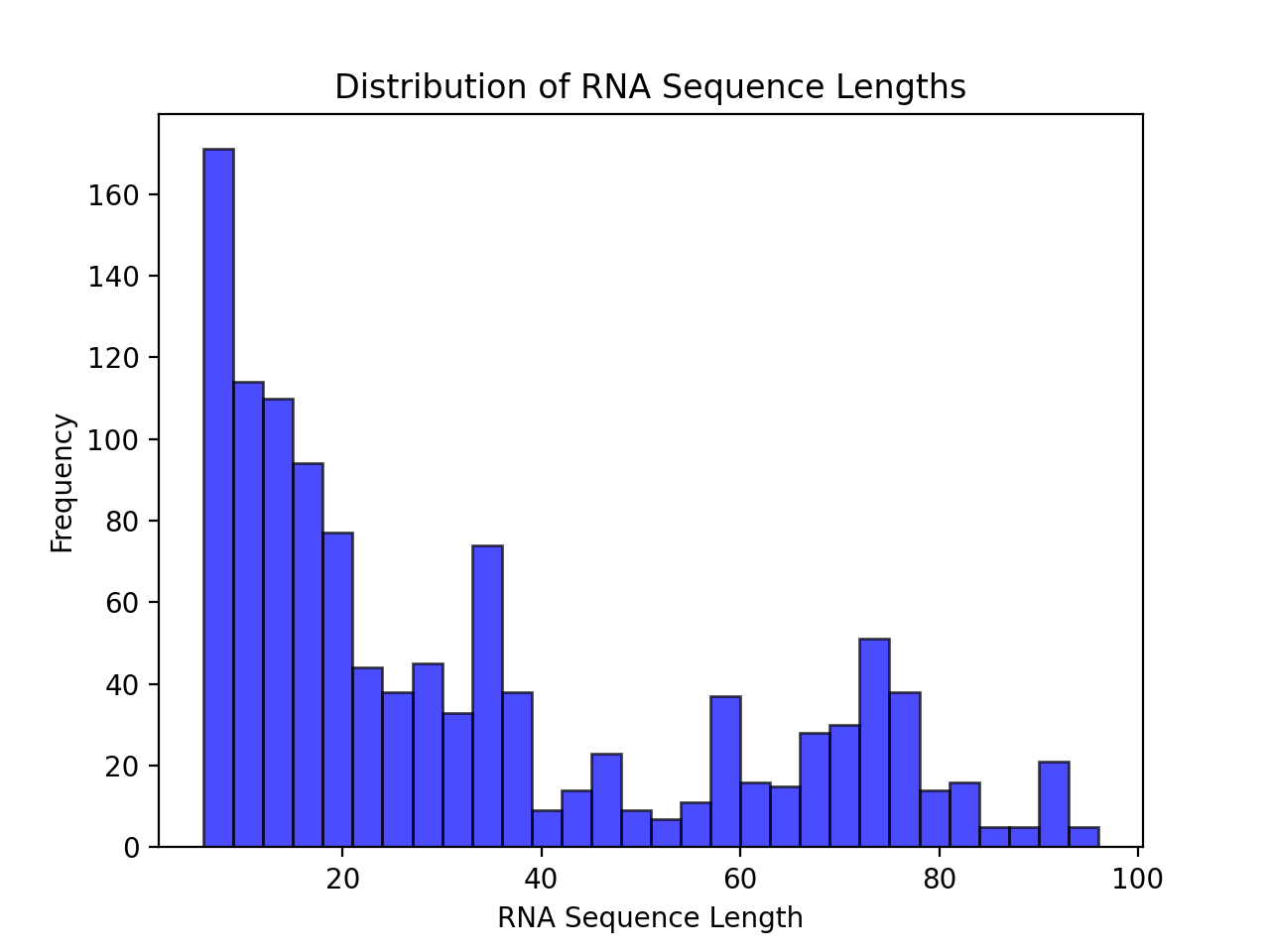}
    \label{fig:lengths}
    \caption{Distribution of RNA lengths in processed PDBBind dataset.}
    \centering
    \includegraphics[width=0.6\columnwidth]{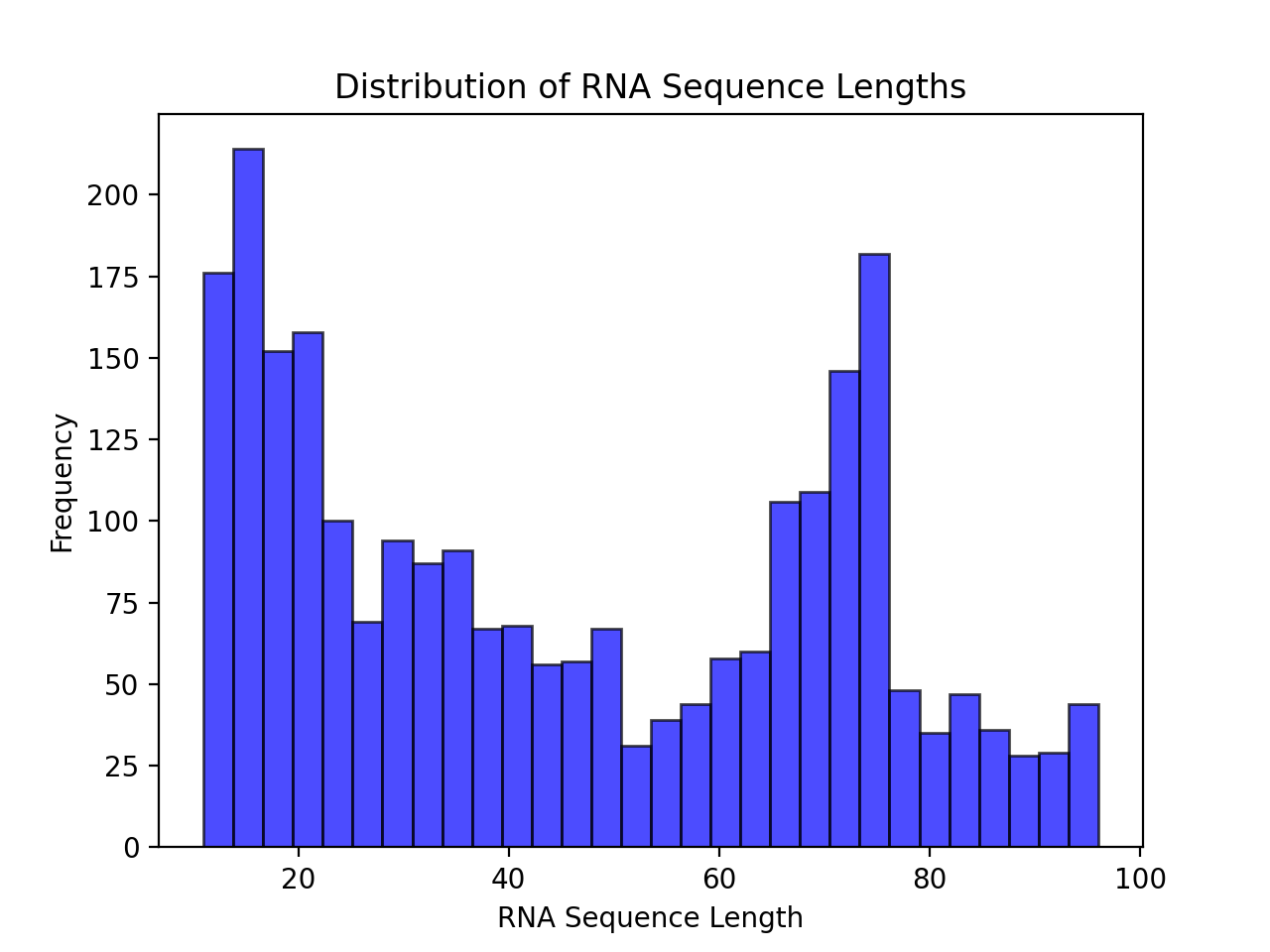}
    \label{fig:lengths2}
    \caption{Distribution of RNA lengths in processed RNASolo dataset.}
\end{figure*}

\section{Dataset Details}

\subsection{PDBBind}

We filter PDBBind to complexes where at least one protein $C_\alpha$ atom and RNA $C_{4}'$ atom were  within $7$ \AA, a threshold that has been used to perform alanine scans for protein-RNA interaction site analysis \cite{kruger2018protein}. For complexes containing many protein-RNA interaction sites, we use the interaction with least distance between the protein $C_\alpha$ atom and RNA $C_{4}'$ atom. We filter to RNA chains of length $\geq 6$ and $\leq 96$, and protein chains are contiguously cropped to length $50$. As shown in Figure \hyperref[fig:lengths]{5}, while there are many examples of short RNAs, the model is also exposed to several longer RNA examples. 

In the sequence similarity split, there are $1015$ complexes in train, $105$ in validation, and $72$ in test. The RF2NA split has $1059$ complexes in train, $117$ in validation, and $16$ in test.

\subsection{RNASolo}

As mentioned, we use the RNASolo dataset for Traj-to-Seq training. The dataset consists of RNAs from apo RNA structures, protein-RNA complexes, and RNA-DNA complexes. As shown in Figure \hyperref[fig:lengths2]{6}, the distribution of RNA lengths in the dataset is somewhat bimodal, with many sequences around length $10$ and around length $75$. In the sequence similarity split, there are $2314$ distinct RNA sequences in train, $106$ in validation, and $78$ in test. In the RF2NA split, there are $2370$ distinct RNA sequences in train, $110$ in validation, and $16$ in test. 

\section{Training and Architecture Details}

\subsection{Noise-to-Seq}

For both splits, the Noise-to-Seq encoder and decoder GVP layers use a node scalar feature dimension of $128$, node vector feature dimension of $16$, edge scalar feature dimension of $32$, and edge vector feature dimension of $1$. On both splits, the model was pre-trained for $100$ epochs using an Adam optimizer with a learning rate of $0.001$, which takes a few hours on an NVIDIA A5000-24GB GPU.

\subsection{Traj-to-Seq}

For both splits, the Traj-to-Seq encoder and decoder GVP layers use a node scalar feature dimension of $128$, node vector feature dimension of $16$, edge scalar feature dimension of $64$, and edge vector feature dimension of $4$. The model was trained for $100$ epochs by an Adam optimizer with a learning rate of $0.001$.

\subsection{RNAFlow}

We fine-tune RNAFlow for $100$ epochs which takes one day on an NVIDIA A5000-24GB GPU, and we use the Adam optimizer with a learning rate of $0.001$.

\subsection{LSTM}

For both splits, the encoder's hidden dimension is $128$ which was selected by a hyperparameter sweep, and the model was trained by an Adam optimizer with a learning rate of $0.001$.

\subsection{Output Rescoring Model}

Following Noise-to-Seq, the encoder and decoder GVP layers use a node scalar feature dimension of $128$, node vector feature dimension of $16$, edge scalar feature dimension of $32$, and edge vector feature dimension of $1$. The node output feature dimension is $256$, and the final feedforward network consists of three fully connected layers with ReLU activation. The model was trained for $50$ epochs using an Adam optimizer with a learning rate of $0.01$.


\end{document}